\documentclass[journal]{IEEEtran}
\usepackage{amsmath}
\usepackage{amsfonts}
\usepackage{amssymb}
\usepackage{graphicx}
\usepackage{cite}
\usepackage{url}
\usepackage{float}

\begin{document}

\noindent
© 2023 IEEE. This is the author-accepted manuscript. DOI: 10.1109/TED.2023.3315250

\vspace{0.5cm}

\title{Fabrication and Experimental Study of Prototype NdFeB Helical Undulators}

\author{Nezah~Balal,
        Eberhard~Bamberg,
        Vladimir~L.~Bratman,
        and~Eyal~Magory
\thanks{Nezah Balal, Vladimir L. Bratman, and Eyal Magory are with the Department of Electrical and Electronic Engineering and the Schlesinger Family Center for Compact Accelerators, Radiation Sources and Applications (FEL), Ariel University, Ariel 40700, Israel (e-mail: nezahb@ariel.ac.il).}
\thanks{Eberhard Bamberg is with Viteris Technologies, LLC, Salt Lake City, UT 84104 USA.}}


\maketitle

\begin{abstract}
An experiment with short prototypes of helical undulators, comprised of longitudinally magnetized helices made from a single piece of a rare-earth magnet, is described. Wire electrical discharge machining (WEDM) in combination with a flat tool and the rotary movement of the workpiece made it possible to achieve high precision in the manufacture of NdFeB helices. An assembly of two oppositely longitudinally magnetized helices with a period of 20 mm and a relatively large inner diameter of 8 mm creates a field of 0.53 T on the axis of the system, which ensures the value of the undulator parameter K close to unity. According to the calculation, Halbach-type helical micro-undulators with periods of (3--6) mm of four helices can provide a field of 1 T and K = 0.28--0.6 at the axis.
\end{abstract}


\section{Introduction}
\IEEEPARstart{H}{IGH-FIELD} undulators and micro-undulators are one of the key elements in FELs of various frequency ranges. The most common are planar Halbach undulators [1], which consist of two parallel lattices of many rare earth bar magnets with a periodic wave-like change in the direction of magnetization. Considering the inevitable spread in the parameters of individual magnets, the assembly and alignment of such structures is a rather difficult task, especially for short undulator periods. To simplify this task, it was proposed in [2] and [3] to use undulators assembled from structures, each of which is made from one rare earth magnet and contains many periods with the same magnetization. In [2], it was suggested to use a planar Halbach lattice assembled from solid combs consisting of equally magnetized elements. In [3], we proposed helical undulators, consisting of helices, each of which is made from a single piece of magnet.

\begin{figure}[H]
\centering
\includegraphics[width=3.5in]{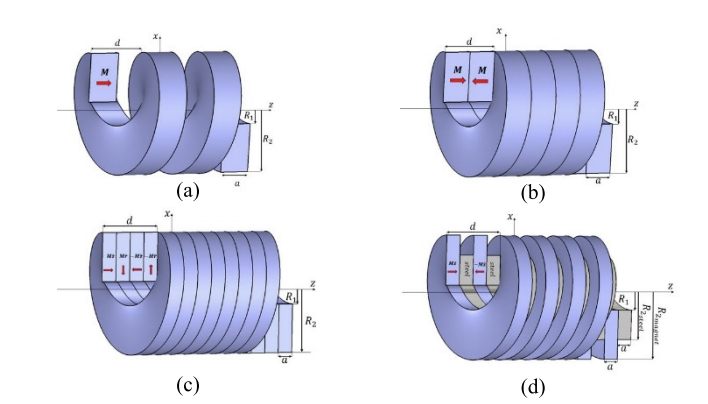}
\caption{Permanent helical undulators with magnetized rare-earth helices. (a) Longitudinally magnetized helix with a rectangular cross section. (b) Two oppositely longitudinally magnetized and half-period shifted helices. (c) Halbach-type undulator of four alternating longitudinally and radially magnetized helices. (d) Hybrid undulator of two longitudinally premagnetized rare-earth and two non-premagnetized steel helices.}
\label{fig1}
\end{figure}

Compared to a planar system with the same electron transport gap, a helical undulator provides a higher field amplitude, and not in one, but in both transverse polarizations. This leads to a significantly higher root-mean-square oscillatory velocity, as well as to particle focusing along both transverse coordinates. Therefore, such undulators can provide a higher radiation efficiency at a shorter FEL length [4], which is especially important for compact X-ray systems.

For the fabrication of a planar micro-undulator from magnetized NdFeB combs, it was proposed in [2] to use the technology of wire electric discharge machining. The same technology makes it possible to manufacture helical undulators and micro-undulators, consisting of magnetized helices with high accuracy. This technology has already been successfully used for the manufacture of helical structures without axial holes for other purposes with periods down to 1 mm [5], [6]. In this article, we report on the fabrication of the first helices specifically for the prototype undulator and the measurement of their fields.

\section{Undulators From Magnetized Helices}

The field of a helix with a fixed magnetization $\vec{M}(\vec{r})$ can be found by integrating the contributions from all the elementary magnetic dipoles $\vec{M}(\vec{r})dV$. In this way, it is easy to analytically calculate the field $\vec{B}_u$ on the axis of an infinitely long and thin uniformly magnetized helix. A simple integration over such helices makes it possible to calculate the field of a thick, azimuthally symmetric helices. Proceeding in this way in the case of a right helix with a constant rectangular cross section and longitudinal (z) or radial (r) magnetization (Fig. 1), we obtain [3]

\begin{align}
\vec{B}_u^z &= B \text{sign} M_z \frac{\sin \eta}{\pi} \int_{\xi_1}^{\xi_2} \xi K_1(\xi) d\xi \left[\hat{x} \sin \zeta + \hat{y} \cos \zeta\right] \label{eq1}\\
\vec{B}_u^r &= B \text{sign} M_r \frac{\sin \eta}{\pi} \int_{\xi_1}^{\xi_2} \xi K_1'(\xi) d\xi \left[-\hat{x} \cos \zeta + \hat{y} \sin \zeta\right] \label{eq2}
\end{align}

Here, $B = \mu_0 M$, $M$ is the remanent magnetization of the ferromagnetic, $\mu_0$ is the magnetic permeability of a vacuum, $\xi_{1,2} = h R_{1,2}$, $\zeta = hz$, $\eta = ha/2$, $h = 2\pi/d$, $d$, $a$ and $R_{1,2}$ are the period, axial width, inner and outer radii of the helix, respectively, and $K_1$ and $K_1'$ are a first-order Macdonald function and its derivative. Equation (1) coincides with that found in [4] by expanding the longitudinal magnetization in a Fourier series and solving the Poisson equation for the surface magnetization on a cylindrical surface $r = R$ with appropriate boundary conditions, namely, the continuity of the magnetic potential and the jump of its normal derivative. This method also ensures that the field is found not only on the helix axis but also in the entire space at once in the form of a series of spatial harmonics.

\begin{figure}[H]
\centering
\includegraphics[width=3.5in]{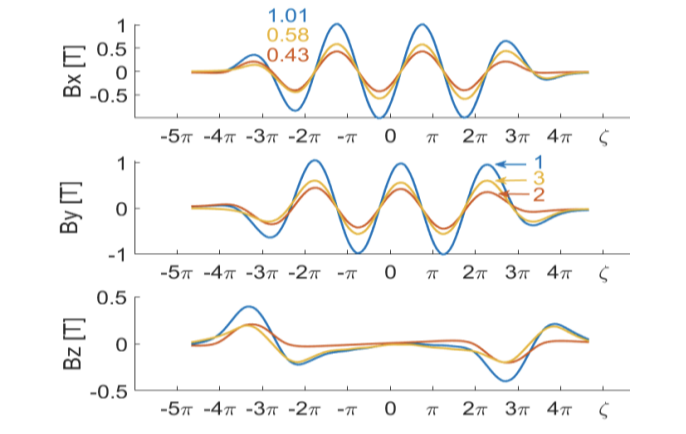}
\caption{Field of a Halbach-type helical undulator comprising four helices with a quarter-period width and alternating axial, and radial magnetizations (1), and contributions from pairs of axially (2) and radially (3) magnetized helices.}
\label{fig2}
\end{figure}

Equations (1) and (2) present the fields of ideal helical undulators. They give good approximations for the transverse fields on the axis even for a small number of periods (5--10) and can be used for estimation even with a shorter length. The field magnitude can obviously be doubled by using the same second helix with opposite axial or radial magnetization and longitudinal displacement of half a period ($h\Delta z = \pi$) relative to the first helix [Fig. 1(b)].

Like for planar undulators, a Halbach-type array of four alternating axially and radially magnetized helices with an axial width equal to a quarter of the period, $a = d/4$, [Fig. 1(c)], provides an additional increase in amplitude (Fig. 2) [3]

\begin{equation}
B = B_r \frac{\sqrt{2}}{\pi} \int_{\xi_1}^{\xi_2} \xi \left[K_1(\xi) + K_1'(\xi)\right] d\xi \label{eq3}
\end{equation}

\begin{figure}[H]
\centering
\includegraphics[width=3.5in]{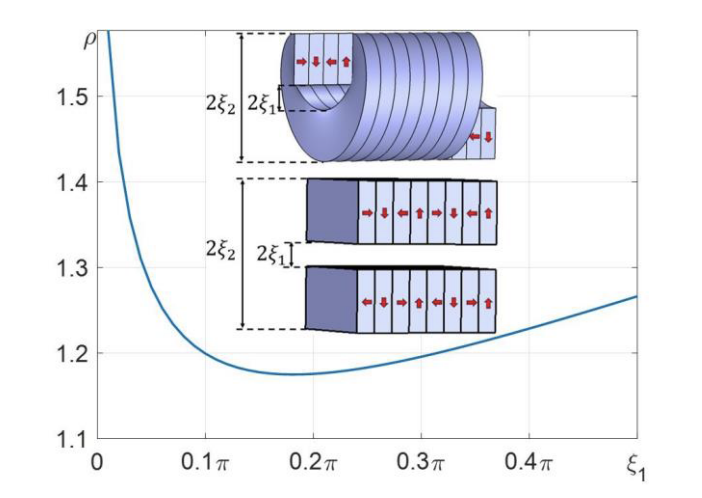}
\caption{Ratio of the field amplitudes on the axis of a Halbach-type helical undulator and Halbach planar undulator with the same gap for electrons $2\xi_1$ and thicknesses $\xi_2 = 2\pi$.}
\label{fig3}
\end{figure}

In a Halbach-type helical undulator, the maximum amplitudes of circularly polarized field components along the x- and y-directions are equal, $B_{x,\text{max}}^{\text{hel}} = B_{y,\text{max}}^{\text{hel}}$. For equal gaps $2\xi_1$ and thicknesses $\xi_2$, these values noticeably exceed the amplitude $B_{x,\text{max}}^{\text{pl}}$ of a linearly polarized component in a Halbach planar undulator even in the case of magnets of infinite bar length along the y-coordinate perpendicular to the direction of magnetization (Figs. 3 and 4). The minimum of the ratio $\rho = B_{x,\text{max}}^{\text{hel}}/B_{x,\text{max}}^{\text{pl}}$ is 1.17 (Fig. 3).

According to formulas (1) and (2), the contribution of radially magnetized helices is greater than that of longitudinally magnetized ones. Fig. 2 shows the distributions of the Cartesian field components of a helical Halbach-type undulator with amplitudes equal to 1 T, and the contributions to this field from axially and radially magnetized helices ($\xi_1 = 1.11$, $\xi_2 = 2\pi$).

\begin{figure}[H]
\centering
\includegraphics[width=3.5in]{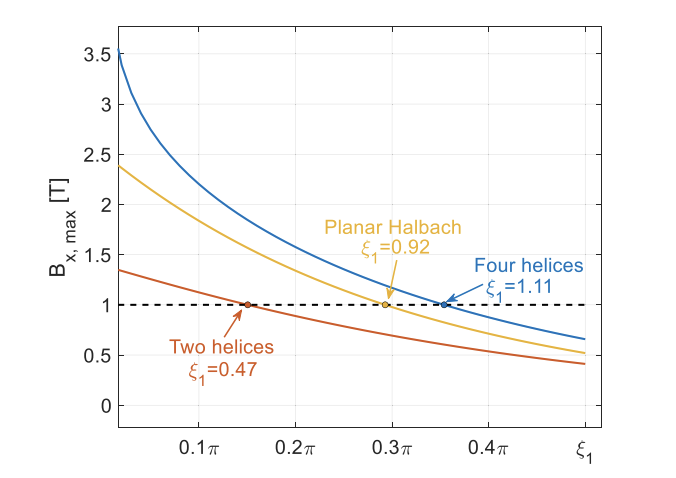}
\caption{Comparison of fields in a Halbach-type helical undulator, a planar Halbach undulator, and an undulator with two axially oppositely magnetized helices ($\xi_2 = 2\pi$). The dimensionless half-gaps are shown, at which amplitudes of 1 T are achieved.}
\label{fig4}
\end{figure}

Obtaining the radial premagnetization of the helices is obviously not an easy task, but it can be solved using a hybrid design [Fig. 1(d)] of two longitudinally premagnetized rare-earth and two non-premagnetized steel (or permendur-vanadium) helices. Like in the corresponding planar hybrid undulators, due to the high magnetic permeability, the field near the inner surface of the non-premagnetized helix, placed in the field of magnetized helices, is directed almost perpendicular to the surface, i.e., nearly radial. According to simulations based on CST Microwave Studio [7], the field of an optimized hybrid undulator on its axis is very close to the field of a Halbach-type helical undulator.

\section{WEDM of Rare Earth Helices}

Wire electrical discharge machining (WEDM) employs a continuously moving conductive wire as an electrode, which serves to cut and shape materials (Fig. 5). The process relies on electrical discharges, or sparks, to remove material from a workpiece. The wire, typically composed of brass or stratified copper, acts as a cathode, while the workpiece to be machined serves as the anode. The gap between the wire and the workpiece is filled with a dielectric fluid, which insulates against electrical discharge until the voltage is sufficiently high. Once the threshold voltage is exceeded, the dielectric fluid breaks down, and an electrical discharge ensues, generating intense heat that vaporizes a minute portion of the workpiece. As the wire moves, the eroded path forms the desired shape in the workpiece.

\begin{figure}[H]
\centering
\includegraphics[width=3.5in]{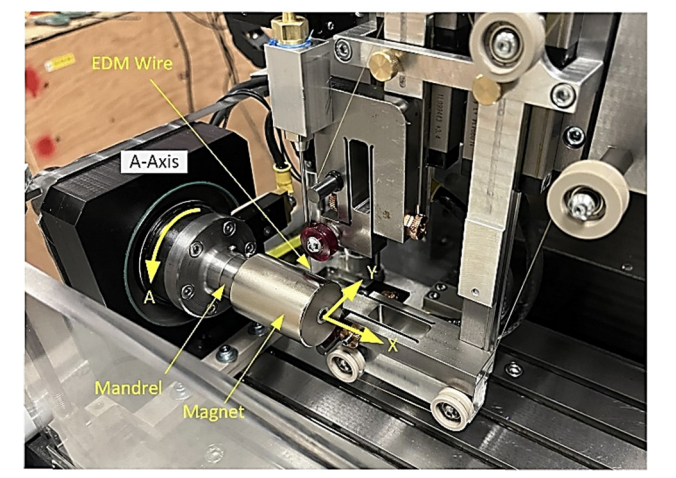}
\caption{Wire electric discharge machining.}
\label{fig5}
\end{figure}

One of the key benefits of WEDM is its ability to manufacture complex parts with high precision and slight damage to the material due to mechanical impact. It is particularly suitable for materials that are difficult to machine using conventional methods, such as hard and brittle substances. This makes it very suitable for cutting rare-earth helices for undulators.

In manufacturing helices, WEDM is combined with a flat tool and rotary workpiece motion (along the x-, y-axes and around the A-axis in Fig. 5, respectively). The flat tool is responsible for lateral movement, enabling the creation of intricate designs with high precision. Simultaneously, the rotary workpiece motion imparts a spiral trajectory to the cutting tool, giving rise to the helical shape of the magnets (Fig. 6).

\begin{figure}[H]
\centering
\includegraphics[width=3.5in]{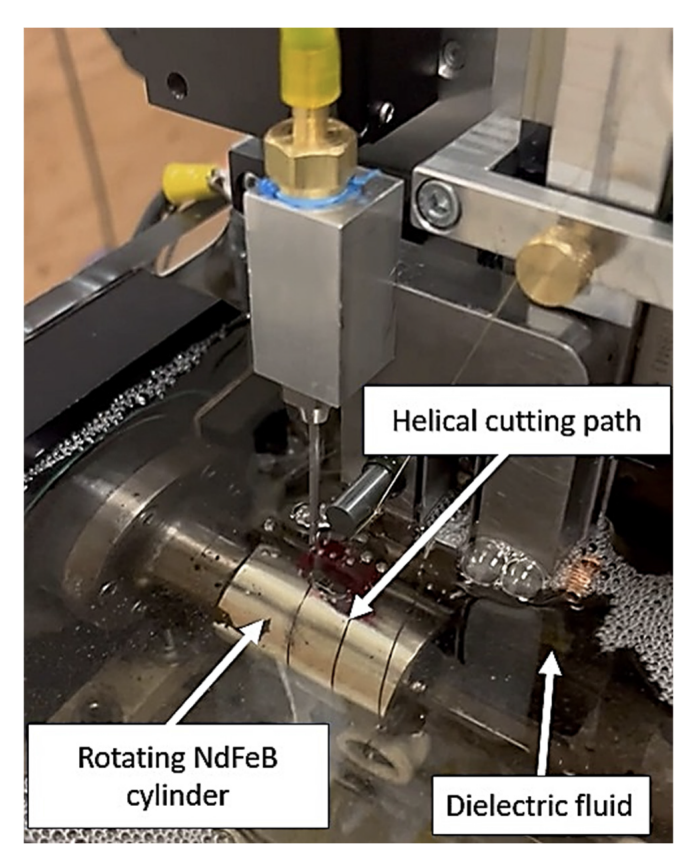}
\caption{Fabrication of a NdFeB helix from a cylinder.}
\label{fig6}
\end{figure}

The integration of WEDM with flat tools and the rotary movement of the workpiece also provides high flexibility and control over the machining. In particular, it allows precise production of NdFeB helices with a period down to 1 mm or even less and material damage only in a very thin micrometer-sized layer [5], [6].

WEDM allows one to fabricate helices both from non-magnetized and magnetized rare earth blanks. To make helices for the prototype undulator we used ready-made non-magnetized NdFeB type N50 cylinders with a strong remanent field $B = 1.4$ T, inner and outer radii of 16 and 4 mm, respectively, and a length of 40 mm. The relatively large inner radius ($\xi_1 = 0.4\pi$) was related to the size of the probe used for measuring the inner magnetic field of the undulator. When using WEDM, two short identical helices with a period of 20 mm and inner and outer radii of 4 and 16 mm were simultaneously cut from the cylinder (Fig. 7).

\begin{figure}[H]
\centering
\includegraphics[width=3.5in]{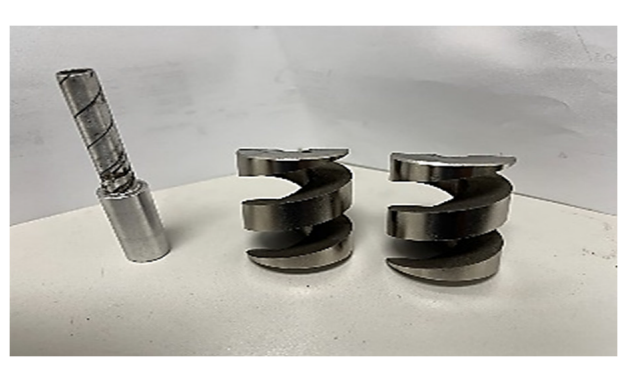}
\caption{Two identical helices obtained simultaneously as a result of a WEDM thin spiral cut of a NdFeB cylinder. On the left is an aluminum support with a spiral track from cutting.}
\label{fig7}
\end{figure}

\section{Magnetization and Assembly of Helices. Repulsion Force}

The fabricated helices were longitudinally magnetized in the field of a pulsed solenoid. Several pulses with a duration of 2 ms and a field greater than 2 T were enough to rich saturation of helices' magnetization. Each such magnetized helix presents the simplest prototype undulator with a fairly strong field [Fig. 1(a)]. To double the magnitude of the helical field, two helices with opposite longitudinal magnetizations, shifted by half a period [Fig. 1(b)], were used.

Two identical non-magnetized helices are easily screwed into each other, forming an almost original cylinder, but with a very thin through helical slot. At the same time, such an assembly is not easy when the helices are magnetized. According to the CST simulation, at zero distance, $z = 0$ between fabricated helices with opposite magnetizations they repel each other with a force of 96 N (Fig. 8). When screwing one helix about half of its length into another ($z = -17$ mm), the force increases to a maximum value of about 160 N, after which it decreases to zero when the movable helix is completely screwed inward. With further movement of the helix, the direction of the force changes sign and the force curve continues symmetrically (Fig. 8).

\begin{figure}[H]
\centering
\includegraphics[width=3.5in]{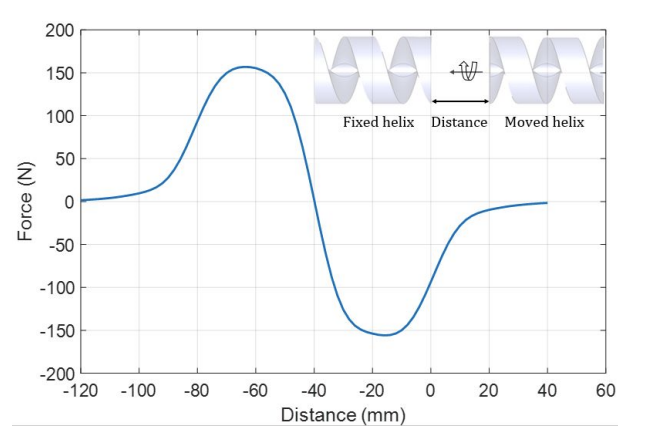}
\caption{CST simulations of force-distance dynamics during assembly of a helical undulator of two oppositely longitudinally magnetized helices.}
\label{fig8}
\end{figure}

It is important that with the increase in the number of periods, the repulsive force changes little, increasing to a value of approximately 103 N at zero distance (Fig. 9). The repulsive force decreases with decreasing undulator period. The pressure it creates is orders of magnitude less than the mechanical strength of NdFeB magnets for micro-undulators with a field on the axis of 1 T and periods in the entire millimeter range. However, when assembling them, it is necessary to prevent the collision of the helices.

\begin{figure}[H]
\centering
\includegraphics[width=3.5in]{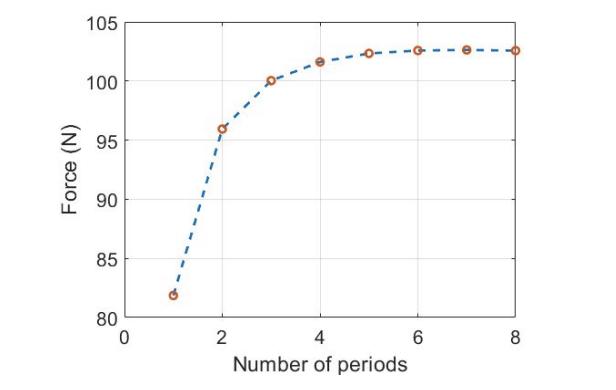}
\caption{Repulsive force at zero distance between helices as a function of the number of undulator periods.}
\label{fig9}
\end{figure}

In the experimental assembly of the undulator, to reduce the risk of impact damage, the helices were located on the central stain-less rod, which excluded their off-axis movement [Fig. 10(a)]. One of the helices was fixed, and the second was gently pushed and twisted by hand [Fig. 10(b)] to obtain a single assembly [Fig. 10(c)].

\begin{figure}[H]
\centering
\includegraphics[width=3.5in]{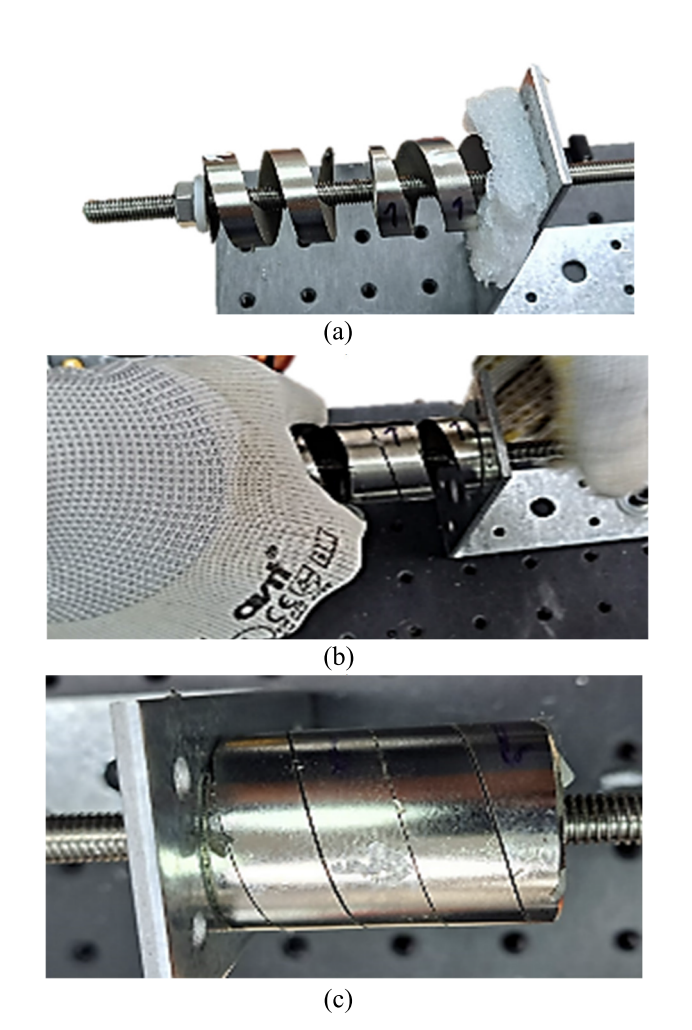}
\caption{Assembling a prototype undulator from two oppositely axially magnetized and repulsive helices. (a) Stringing the helices on the central rod. (b) Pushing and screwing the helices into each other. (c) Prototype undulator.}
\label{fig10}
\end{figure}

\section{Comparison of Calculated and Measured Fields}

According to (1), an infinite helix with the remanent field $B = 1.4$ T, longitudinal magnetization, and parameters indicated in Section III creates a field $B_u = 0.261$ T on its axis. A close value $B_u = 0.258$ T is obtained from CST simulations. The measured fields of each helix are also very close to these values (Fig. 11).

\begin{figure}[H]
\centering
\includegraphics[width=3.5in]{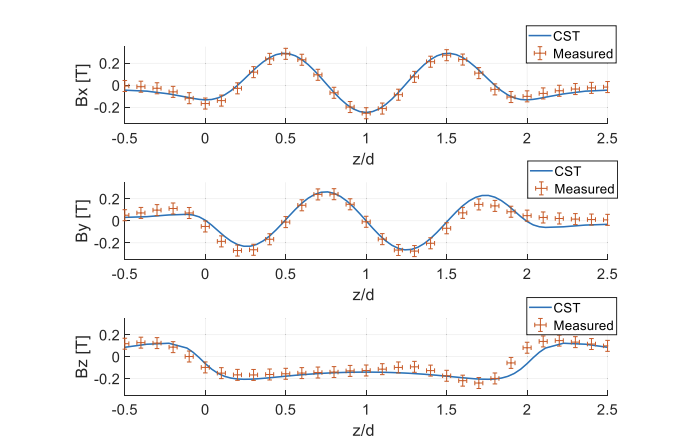}
\caption{CST simulations and measurements of the transverse field on the axis of a longitudinally magnetized helix with a period of 20 mm and an inner radius of 4 mm.}
\label{fig11}
\end{figure}

The connection of two helices with opposite magnetization and axial shift by half a period gave us a field of 0.53 T. The measured field agrees very well with this analytical value and with the CST simulation results (Fig. 12).

\begin{figure}[H]
\centering
\includegraphics[width=3.5in]{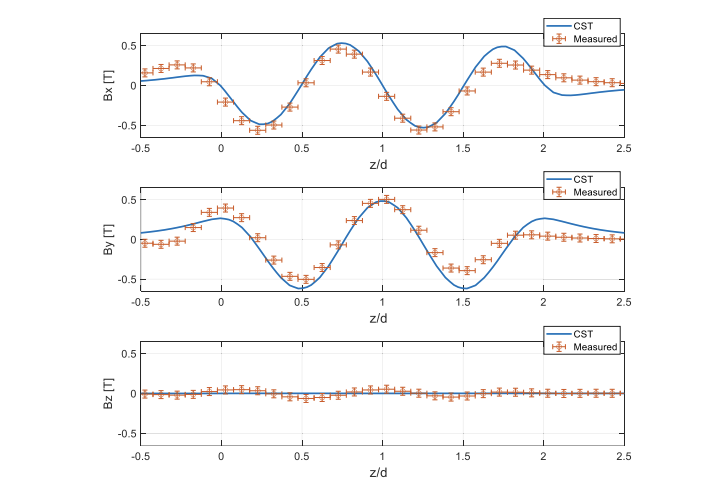}
\caption{CST simulations and measurements of the transverse field on the axis of a prototype undulator consisting of two opposite-axial magnetized and half-period shifted helices with periods of 20 mm and inner radii of 4 mm.}
\label{fig12}
\end{figure}

\section{Prospects for the Use of Magnetized Helices in THz and X-Ray FELs}

In the previous sections, the efficiency of WEDM and magnetization in the field of a pulsed solenoid was confirmed to create high-quality high-field NdFeB helices. The calculated values of the magnetic forces and the developed method for assembling the helices inspire optimism regarding the development of new types of permanent helical undulators and micro-undulators for various applications. Such undulators with centimeter periods and an undulator parameter $K \geq 1$ can be effectively used in high-power FEL sources of THz radiation [8], [9], [10], [11]. Halbach-type micro-undulators of magnetized helices with periods down to (3--6) mm and corresponding channel diameters for electrons of the order of (1 and 2) mm, as well as planar micro-undulators with the similar gap [12], can provide fields on the axis of the order of 1 T (Figs. 2--4), but in both transverse directions. The use of such micro-undulators can provide a significant increase in the efficiency of compact X-ray FELs.

\end{document}